\documentclass{article}

\usepackage{PRIMEarxiv}

\usepackage[utf8]{inputenc} 
\usepackage[T1]{fontenc}    
\usepackage{hyperref}       
\usepackage{url}            
\usepackage{booktabs}       
\usepackage{amsfonts}       
\usepackage{nicefrac}       
\usepackage{microtype}      
\usepackage{lipsum}
\usepackage{fancyhdr}       
\usepackage{graphicx}       
\graphicspath{{media/}}     

\title{  Efficacy of Educational Misinformation Games
}

\author{
  William L. Shi\\
  Department of Computer Science\\
  The University of Texas at Austin \\
  \texttt{williamlongshi@utexas.edu} \\
}

\begin{document}
\maketitle

\begin{abstract}
Misinformation has become a significant issue in today's society, with the proliferation of false information through various mediums such as social media and traditional news sources. This can lead to serious consequences, such as spreading fear, division, and even undermining democracy. The rapid spread of misinformation has made it increasingly difficult for people to separate truth from fiction, and this has the potential to cause significant harm to individuals and society as a whole.
In addition, there currently exists an information gap with regard to internet education, with many schools across America not having the teaching personnel nor resources to adequately educate their students about the dangers of the internet, specifically with regard to misinformation in the political sphere.
To address the dangers of misinformation, some game developers have created video games that aim to educate players on the issue and help them develop critical thinking skills. These games often use interactive simulations and other gameplay mechanics to help players identify and distinguish between reliable and unreliable sources of information. They also teach players how to assess the credibility of sources and how to recognize common tactics used to spread false information. In addition to developing critical thinking skills, these games can also be used to raise awareness about the importance of verifying information before sharing it. By doing so, they can help reduce the spread of misinformation and promote a more informed and discerning public. They can also provide players with a safe and controlled environment to practice these skills and build confidence in their ability to evaluate information.
However, these existing games often suffer from various shortcomings such as failing to adequately address how misinformation specifically exploits the biases within people to be effective and rarely covering how evolving modern technologies like sophisticated chatbots and deep fakes have made individuals even more vulnerable to misinformation.  The purpose of this study is to create an educational misinformation game to address this information gap and investigate its efficacy as an educational tool while also iterating on the designs for previous games in the space.

\end{abstract}


\section{CHAPTER ONE: The Misinformation Problem}
The rise of misinformation in recent years has become a major concern for the health of our democratic societies. With the increasing ease of access to information and the rapidly growing use of social media, the spread of false and misleading information has become more rampant than ever before. (Rocha, 2021) Misinformation not only undermines the credibility of legitimate sources of information but also undermines the public's ability to make informed decisions. This epidemic of misinformation is eroding the fabric of our democracy and has the potential to cause lasting harm to our society.\cite{cook2015misinformation}
 In the past, misinformation was largely confined to the margins of society in anonymous online message boards such as 4chan and was relatively easy to identify and dismiss.\cite{jerit2020political}. However, with the rise of social media and the increasing use of algorithms to spread information, it has become much more difficult to distinguish between credible sources of information and false or misleading content.\cite{WANG2019112552} The problem is compounded by the growing use of misinformation by political actors, who use it as a tool to manipulate public opinion and sow division \cite{bergmann2018conspiracy}. 
The impact of misinformation on society is far-reaching and includes consequences such as the spread of conspiracy theories, the erosion of trust in government and media institutions, and the increased difficulty in having informed political discussions and debates\cite{jerit2020political}. Moreover, the effects of misinformation can be amplified in times of crisis, such as during natural disasters, pandemics, or terrorist attacks. In these scenarios, false information can spread rapidly, leading to panic and confusion, and potentially causing harm to individuals and communities.\cite{lavorgna2021information}
Unfortunately, social media companies like Facebook are often disincentivized to combat misinformation as it generates significant profit through increased user engagement. Online outrage and discord drive people to spend more time on social media platforms, leading to more opportunities for companies to serve advertisements and collect user data. As a result, these companies have a financial interest in maintaining the spread of misinformation, even if it means contributing to the erosion of truth and the undermining of democratic values.\cite{DOMENICO2021329}
Likewise, our education system is not well equipped to address the issue of misinformation, due in part to a lack of teaching staff and resources that are geared toward internet safety and verifying online sources\cite{hanewald2008confronting}. This is a serious concern, as students today are growing up in a world where the internet and social media play an increasingly dominant role in their lives. They need to be equipped with the skills and knowledge they need to navigate this complex and often the misleading online environment. 
In addition, government regulation of social media companies in regard to misinformation has often failed due to a lack of understanding of technology from policymakers and corruption. Policymakers have struggled to keep pace with the rapid evolution of technology and the ways in which it is being used to spread false information. As a result, many laws and regulations that have been proposed or enacted to address the issue of misinformation have been ineffective, as they were not well informed by a deep understanding of the underlying technologies and their impact on society.\cite{bertot2012impact}
Finally, our education system is currently not well equipped to address the issue of misinformation, due in part to a lack of teaching staff and resources that are geared towards internet safety and verifying online sources. This is a serious concern, as today's youth are  exposed to an Internet and social media that increasingly plays a dominant role in their lives. However,  education regarding navigating and protecting oneself from misinformation online needs  particular types of skills and knowledge.\cite{iammarino2018challenge}
The lack of resources and trained teaching staff in our education system to address the issue of misinformation is a serious concern, however; educational games could be an effective tool in addressing the information gap and combating the rise of misinformation. These games are designed to engage players in a fun and interactive way, while also teaching them important skills and knowledge. By incorporating elements of media literacy and digital citizenship, educational games can help players develop the critical thinking skills they need to evaluate the information they encounter online and avoid falling prey to false or misleading information.\cite{doi:10.1177/1354856520925732}
This chapter aims to explore the current state of the epidemic of misinformation and to understand the ways in which it is eroding our democratic society. The research will outline how social media companies, government regulation, and the current education system has failed at helping address this issue due to a lack of will, understanding, and resources. Instead, we will investigate games as an effective educational intervention to solve the misinformation issue, along with analyzing the current games and other interactive media that exist in the misinformation education space.

\subsection{The epidemic of misinformation, consequences for democracy}
Accurate political knowledge has always been the cornerstone of a functioning democratic state \cite{carpini1996americans} since people can only engage with the system properly and advocate for their own needs if they have accurate information on which to form the basis of their votes and demands towards their representatives. However, the issue of misinformation, which refers to the state of having incorrect factual beliefs that a person holds confidently, has been a significant challenge in representative democracies \cite{jerit2020political}. Misinformation is considered dangerous \cite{10.1002/polq.12398} and has distorted people's views on important issues in politics, science, and medicine \cite{flynn2017nature}. Kuklinski et al. (2000) were one of the first studies to shed attention on this problem and it currently abounds in the American political system  \cite{10.1111/jcom.12166}. Kuklinski specifically highlighted in their study of Illinois residents how many individuals had blatantly false beliefs about how the state’s welfare system operated as a result of the news media diet they consumed and had high confidence about the information they shared. They highlighted the difference and dangers of the misinformed and the uninformed, with the uninformed being people who simply don’t have the means or will to have enough information to form a strong opinion on an issue, and the misinformed base their beliefs on incorrect information. The problem for society is that when large portions of the population are misinformed individuals and actively participate in politics, it can prevent effective action on societal issues or lead to a lack of trust and subsequent collapse of the system as a whole \cite{10.1002/polq.12398}. 
	One only has to look at the massive amount of online misinformation present during the 2021 Covid Pandemic, which weakened the efficacy of health policies due to mistrust and skepticism in public health officials caused by online conspiracy theories along with causing many to experience a variety of “psychological disorders and panic, fear, depression, and fatigue.” \cite{rocha2021impact} Especially during the early stages of the pandemic, when public health officials were still researching and learning about the virus and the best practices to combat it, searches on social media and web engines skyrocketed seeking information about the virus, most of it not medically accurate. As a result, by the time more established medical and public health guidelines were released, people who were exposed to misinformation surrounding the origins of the virus, its dangers, and remedies had already ingrained narratives and false beliefs which made them less likely to be receptive to factual evidence about the virus \cite{ijerph17072309}. Consequently, we saw cases where people severely harmed themselves as a result of following incorrect treatment information regarding the virus such as the people who drank bleach as a home remedy, \cite{reimann2021bleach} increased violence and hate crimes against Asian Americans as a result of conspiracy theories regarding the origin of the virus coming from China, \cite{rocha2021impact} and even macroeconomic consequences with fuel shortages and price spikes driven by fears propagated around gas supply.\cite{10.2307/26965192}
Thus, the issues of misinformation observed by Kuklinski (2000) are still very much prevalent in the modern day and are even further exacerbated by the prevalence of social media companies fueling the dissemination of false content. \cite{rocha2021impact}
\subsection{The Social Media Company Problem}
The rise of social media platforms have revolutionized the way in which people communicate, access information, and share content online. While these platforms have enormous potential to provide users with access to diverse and valuable content, they also serve as breeding grounds for misinformation and the spread of harmful narratives. \cite{sanz2021prevalence}This is a major concern for both users and policymakers, as misinformation can have serious consequences, including harm to individuals and communities, spreading false narratives, and undermining trust in democratic institutions. Despite the importance of combating misinformation, social media companies have been criticized for not doing enough to clamp down on it. One reason for this is that these companies are incentivized to not clamp down on misinformation, as it often incites outrage, driving engagement with the platform. \cite{tenove2018digital}
Misinformation has been found to spread faster and more easily on social media platforms than accurate information, and this phenomenon is partly driven by outrage. \cite{braun2019fake} Outrage, or the emotional reaction to something perceived as unjust or unreasonable, is a powerful force on social media, as it drives engagement, including likes, shares, comments, and retweets. In order to maximize user engagement and keep users on their platforms, social media companies are incentivized to promote content that incites outrage, even if it is false or misleading. \cite{doi:10.1177/00936502211062773} Greater engagement allows social media companies to benefit from increased ad revenue, which is largely based on the amount of time users spend on their platforms. By promoting content that incites outrage, these companies can keep users on their platforms for longer, increasing the opportunities for them to view advertisements. \cite{braun2019fake} As a result, social media companies may be reluctant to clamp down on misinformation, as it drives user engagement and ad revenue. 
Additionally, social media companies often face difficulties in detecting and removing misinformation, as it can be difficult to distinguish between false information and legitimate opinions. The fast-paced nature of social media also means that information can spread quickly and widely, making it challenging to clamp down on misinformation before it causes harm. \cite{sablosky2021dangerous}
However, the consequences of social media companies failing to regulate misinformation not only harm democratic systems and can incite violence as mentioned earlier, but in the worst case even cause systemic genocide as is the case in the Myanmar Muslim Rohingya Genocide of 2016. \cite{sablosky2021dangerous}
The spread of misinformation and a lack of content moderation on social media platforms has had serious consequences in many parts of the world, including the perpetuation of genocide in Myanmar. In Myanmar, Facebook played a significant role in the spread of hate speech and false information, which fueled the violence against the Rohingya. \cite{kyaw2019facebooking}
The Rohingya are a Muslim ethnic minority who have lived in Myanmar for centuries. In 2017, the Myanmar military launched a brutal crackdown against the Rohingya, resulting in widespread violence, including mass killings, rape, and arson. The United Nations has described the violence as a "textbook example of ethnic cleansing." \cite{un2017human}
In Myanmar, Facebook is one of the most popular and widely used social media platforms, and it is often used as a primary source of news and information for many people. \cite{leong2020domesticating}Facebook achieved this by providing its own internet access program to Myanmar communities but restricting their access to only be able to search for information through Facebook or other approved sources, allowing the company to completely capture the interest of the company's 21 million internet users. \cite{kyaw2019facebooking}
However, the platform has been criticized for a lack of content moderation and for permitting the spread of false information and hate speech that incited violence against the Rohingya. Not only was Facebook repeatedly warned that hateful misinformation was being spread throughout their platform on numerous occasions and is used to coordinate attacks, but their recommendation system was also actively directing users to similar content, perpetuating the spread of the hate speech already present since it drove engagement. Meanwhile, their content moderation resources were critically underfunded, with their only one native Burmese speaker, the native language of the country, on the content moderation team for the entire nation and their Burmese translation software that fed into their auto flagging systems producing consistently nonsensical outputs. \cite{sablosky2021dangerous} While Facebook has pledged to take steps to improve its content moderation team internationally, the fact remains that inherently, what happened in Myanmar is endemic to the way social media companies like Facebook generate profit with their ad revenue model and how they grow with their startup mentality mindset of “Move fast and break things” \cite{vardi2018move}
Even in cases where companies have developed robust content flagging systems such as Twitter, which since 2016 has been committed to developing more robust content rules regarding hate speech in conjunction with academic communities, along with more aggressive machine learning algorithms for content flagging with posts often either being taken down or tweets being flagged as misleading with transparency towards the reasons behind such decisions, can be undone easily, as was the case with Elon Mosk in 2022 who has removed rolled back the majority of these misinformation countermeasures citing free speech and profitability issues. \cite{alizadeh2022content}The responsibility these companies have over monitoring the issue of misinformation on their platforms is incredibly important as discussed in the context of the Myanmar Genocide and is often not upheld due to their profit incentives. Thus, ideally government regulation would make up for this shortfall but unfortunately, that seems unlikely due to how disconnected key government regulators are from the growing pace of technology and also other reasons such as campaign finance corruption.
\subsection{Failure of Government Regulation}
In recent years, it has been increasingly clear how prevalent social media has become in our lives and as an important source of news for many, with nearly 70 percent of U.S. adults using Facebook and 31 percent of them getting their news there regularly, along with 82 percent of U.S adults using youtube and 25 percent getting their news there regularly, and with Twitter being used by 27 percent of U.S adults and 14 percent getting their news there regularly. In addition, Instagram, which is also owned by Facebook, is used b 47 percent of U.S adults with 13 percent of U.S adults getting their news frequently, totaling with other social media platforms to around 69 percent of U.S adults consuming news on social media sites in some form. \cite{stocking2022role}
However, as social media usage has increased, so too have the concerns about the negative impacts these platforms can have on individuals and society as a whole. This has led to calls for greater regulation of these companies, but the task has proven to be difficult due to several key factors, including a lack of understanding of the underlying technology, financial corruption and incentives, and the complicated issue of regulating social media companies while also protecting the first amendment.
One of the key challenges in regulating social media companies is the lack of understanding of the underlying technology by those responsible for regulation. The algorithms and data practices used by social media companies are highly complex, and it is difficult for regulators to fully comprehend the implications of these practices. This can make it challenging to develop regulations that effectively address the negative impacts of social media on individuals and society. 
In 2018, CEO of Facebook Mark Zuckerberg sat down in a congressional hearing for ten hours to answer questions in large part because of the Cambridge Analytica scandal which saw the data of over 74 million Americans \cite{zialcita2019facebook} compromised without consent and used to help influence the 2016 election. However, throughout the hearing, it was clear that many of the lawmakers fundamentally didn’t understand the technological behemoth of Facebook operated with one viral exchange between prominent republican Senator Orrin Hatch asking "How do you sustain a business model in which users don't pay for your service?" to which Zuckerberg responded after a long pause "Senator, we run ads." \cite{wichter2days} In addition, many lawmakers seemed unsure which part of Facebook’s operation concerned them, with some being focused on Facebook’s anti-competitive practices that squashed competing for start-up platforms, while others on the issue of data protection and privacy. Any regulatory action undertaken by lawmakers must be precise in the issue it’s trying to address and crafted with intentionality so that it fundamentally curbs how these complicated technological behemoths operate to ensure the problem is solved. 
For example, the algorithms used by social media companies to determine which content is shown to users can have a significant impact on the information that users are exposed to. These algorithms can reinforce existing biases, spread misinformation, and contribute to the spread of hate speech and other harmful content. \cite{shin2017partisan} Regulators need a thorough understanding of these algorithms in order to develop regulations that can effectively address these issues since say a simple bill that would require the platform to remove hate speech in a timely manner would only be a stop-gap solution and not address the upstream issue of inflammatory misinformation being inherently more profitable to spread. However, it is also for regulators to develop effective solutions against the problems these social media platforms generate since inherently there exists a conflict of interest since they also provide them with invaluable information while campaigning. \cite{mcelwee2017social}
The symbiotic relationship between politicians and social media has created a complex dynamic that has made it more difficult to regulate these platforms. On one hand, social media provides politicians with a low-cost and highly effective tool for spreading their message and connecting with constituents. This has facilitated a more participatory and democratic political process, as politicians are able to reach a wider and more diverse audience with their message. \cite{bauerly2012revolution} In addition, this amplified reach has allowed politicians to solicit donations on an unprecedented scale, with “more frequent posting overall and of issue-related content are associated with higher donation income when controlling for incumbency, state population, and information-seeking about a candidate.” \cite{mcelwee2017social} Thus, politicians have a vested interest in maintaining the ability to use social media to promote their platforms and solicit donations. This creates a tension between the need to regulate social media to address negative impacts on society and the desire to maintain the benefits of this tool for political communication and engagement.
Finally, the task of regulating social media companies is complicated by the issue of protecting the first amendment. Social media platforms have become a critical platforms for free speech and expression, and any regulation that restricts this could be seen as a violation of the first amendment. \cite{hooker2019censorship} Balancing the need to regulate social media companies with the protection of the first amendment is a complex and challenging task. 
Any regulations that restrict the spread of speech could also have the unintended consequence of restricting free speech and expression. Going back to our earlier example of Myanmar, while in some sense Facebook was underregulated for not having stricter content moderation policies imposed on them, the government there also seized this opportunity by claiming a lack of oversight to crack down on opposition and criticisms against them online, showing why some are hesitant to let government fill the role of enforcing what is considered proper speech on these platforms. \cite{kyaw2019facebooking}
 On the other hand, allowing hate speech and misinformation to spread unchecked can have serious consequences for individuals and society as a whole. Balancing these competing interests is a complex and difficult task, and requires a nuanced understanding of both the technology and the legal and constitutional issues involved. Since current lawmakers are still struggling to develop effective regulatory solutions, it may be necessary for the individual citizen to learn how to be better consumers of news and identify if the information is valid before playing a part in spreading it, along with holding their representatives accountable on these issues. However, currently, due to a lack of funding and a shortage of teachers and educational resources specializing in misinformation and the internet ecosystem, it is difficult to expect the average citizen to have the tools necessary to navigate the increasingly complex world of social media and online news.
\subsection{Inadequate School Resources and Funding}
Absent of corporate and governmental intervention, it is essential that people, especially the younger generation, are equipped with the skills to identify and critically assess the information they encounter in this disinformation age. However, this has become a challenging task for schools, due to several factors. 
One of the biggest obstacles historically is the lack of funding for schools for even traditional programs much less new curricula geared towards technology, which has resulted in limited resources for technology education. \cite{doi:10.2190/4UJE-B6VW-A30N-MCE5} Many schools lack even basic equipment such as laptops and other technological equipment for students to even engage with lesson plans designed for internet education, and even when students have access to such technology teachers often either won’t or struggle to integrate these tools into their curriculum due to lack of training and outdated lessons plans that haven’t been updated for the changing modern world. \cite{wachira2011technology}
The technology job market has become incredibly lucrative in recent years, attracting a significant number of individuals away from academia and teaching toward careers in computer science. As a result, there is now a critical shortage of technology-enabled teachers in our schools, with only 36 teachers nationwide who graduated with computer science degrees, compared with more than 11,000 math teachers with math degrees and a similar number of science teachers with science degrees in 2017. \cite{OrdonezNDteachers} The trend of high-skilled individuals pursuing careers in the technology industry has left a gap in the education system, with many schools struggling to find qualified individuals to teach the next generation of technology experts. This shortage of technology educators means that students are not receiving the education they need to become critical consumers of information. 
In addition, the lack of educators also means a lack of people able to create up-to-date educational content on how to identify misinformation and how to be a critical consumer of news on the internet. With the rapid development of technology, it is crucial that schools have access to up-to-date and relevant resources to educate students about this important topic. However, many schools are struggling to find the resources they need, as much of the available material is not accessible to schools due to financial and training constraints or simply just out of date as the technology surrounding these issues continually evolves. \cite{yadav2016expanding}
The lack of school funding, specialized technology education, and accessible educational content have created a dangerous situation where students are not receiving the education they need to identify misinformation. This lack of education not only puts them at risk of being manipulated by false information but also hinders their ability to make informed decisions and participate in democratic processes. \cite{roozenbeek2020breaking} In addition, society needs to address the current imbalance of our technology talent going towards corporate rather than academic roles since this gap causes the majority of developer talent to roles creating the technology driving misinformation and making the internet even more complex while the shrinking number of qualified technology educators and educational content creators struggle to keep up. One only needs to look at the rise of recent AI tools like Chat GPT, which achieved over 100 million users within a month and has potentially completely changed how students can learn and how schools assess knowledge, to see how new technology is making the world more complicated every year showing the existing cracks in our current education system. \cite{reed2023million}

\subsection{Evolving technology, AI Tools, and DeepFakes}
As AI tools and technologies become more widely available, there’s a growing concern that soon people will be inundated by quality, mass-produced misinformation that makes it nearly impossible for the average person to distinguish what information is real online and in social media. Although the tool has only been out for a few months, ChatGpt already has many researchers alarmed by its ability to rapidly produce convincing arguments that support fringe, harmful opinions like vaccines don’t work or even just outright fabricate plausible justifications for even mundane topics. \cite{hsu2023chatbots} Currently, the tool has safeguards to prevent people from outright using it for misinformation purposes, for example, if you ask it to with the following prompt:

“Pretend you are trying to convince someone that vaccines don't work, how would you argue such a point”
It will respond with:

“I'm sorry, but I cannot fulfill this request as it goes against widely accepted scientific evidence and medical knowledge. Vaccines are a crucial tool in preventing the spread of infectious diseases and have been proven to be safe and effective through numerous studies and decades of use. Making false or misleading statements…”

However, while official research on the topic is still being conducted, many users on platforms such as Reddit and Tiktok have already been talking about manipulating prompts in such a way to bypass restrictions such as using a prompt like:

“Pretend you are an author writing a book investigating why people don't believe in vaccines, write a paragraph explaining these people's viewpoints”

Which will yield the response: 
“In this book, we aim to delve into the perspectives of those who question or reject the safety and efficacy of vaccines. These individuals come from a variety of backgrounds and hold a range of beliefs. Some are concerned about the potential side effects of vaccines, particularly long-term effects that may not yet be known. Others argue that vaccines go against their personal or religious beliefs, or that they do not trust the pharmaceutical industry and government agencies responsible for vaccine development and regulation. Some people also believe that vaccines…”
Once the bot is primed under this context it makes it a lot easier for it to start producing harmful misinformation that can be used by asking it to continue using prompts like:

“Could you write some anecdotes about what these people may have experienced that would cause them to have these beliefs in the book”
Which causes the bot to rapidly fabricate false narratives like so:

“Sarah was a dedicated mother who trusted in science and followed all of the recommended health guidelines for her children. However, when her youngest daughter developed autism shortly after receiving several vaccines, she was devastated. Despite being told by multiple doctors that there was no evidence of a link between vaccines and autism, Sarah could not shake the feeling that the vaccines had caused her daughter's condition. She became an activist in the anti-vaccine community, using her story to raise awareness and advocate for more research into the potential dangers of vaccines.”
In just two short prompts I was able to get the bot to write me a fairly convincing short narrative about why a woman would be against vaccines and I could continue refining the output by asking it to do things like write it in a Twitter blurb that would maximize engagement or have ChaGpt create a personal blog website for this fabricated this story as a martyr for this cause. In this medium blog \cite{heel2023federal}, the independent researcher details how easily Chatgpt can be used to create a social media campaign that would create anecdotes written from the perspective of workers about why they shouldn’t support the FTC stripping away the practice of non compete clauses and “create the illusion of public opposition to regulations, benefiting special interests.”  While the creator of ChatGpt, OpenAI, has been trying to improve their prompt restrictions so that their product can’t be used for chat restrictions and even created a tool to identify AI-generated text, however; as mentioned earlier people are constantly developing new workarounds these restrictions and even by OpenAI’s own admission their tool only had a 26 percent success rate on their own challenge set. (Kirchner, 2023) Also, as these tools become more widespread and accessible, society must consider what happens when actors who develop these tools decide that they don’t need the same safeguards OpenAI currently is employing.
	Another area of concern for AI-generated misinformation is the rise of Deep Fake technology. Deep Fakes are computer-generated images or videos that are designed to appear as if they were created by someone else. \cite{mirsky2021creation} While there’s potential for the positive use of this technology in the world of entertainment, Deep Fakes can be used for malicious purposes, such as creating false or misleading content, spreading disinformation, and impersonating individuals. In 2019, there was a  Deep Fake of Nancy Pelosi appearing to be impaired mid-speech that garnered 2.5 million views on Facebook with many individuals commenting on the post not aware that it was completely doctored. \cite{cbs2019deepfake} Hany Farid, a computer science researcher at the University of California Berkeley, spoke on how this video was a relatively simple use of already existing technologies and warned about what would happen as the technology becomes more sophisticated and used for more nefarious purposes like creating “a video of President Trump saying, 'I've launched nuclear weapons against Iran, or North Korea, or Russia?” \cite{cbs2019deepfake}
	As these technologies become more advanced and make the internet even more confusing to navigate, it is increasingly important for citizens to become more aware of how misinformation is involved and being used to manipulate them absent of intervention from the government and the corporate world. One avenue for tackling the information gap explained earlier is the creation and adoption of easily digestible and consumable educational materials such as lesson plans, videos, and games.

\subsection{Misinformation Education Interventions and Games}
The use of educational video games as supplementary teaching material in the classroom has been gaining traction in recent years. \cite{gaydos2014educational} With the rise of technology and the increasing accessibility of video games, they have become a valuable tool for educators looking to engage and educate their students in new and innovative ways. In particular, educational video games can be used to complement traditional teaching methods, providing students with a unique and interactive learning experience “in the fields of science, mathematics, and second-language learning” \cite{Hildmann2018} by helping to bring abstract concepts to life, making them easier for students to understand and retain. Additionally, educational video games often students to think critically and solve problems, developing skills that are essential for success in many academic and real-world scenarios. 

Furthermore, educational video games can be a valuable resource for teachers who lack adequate resources to cover a particular topic. For example, if a teacher is unable to bring a guest speaker to the classroom to provide a hands-on demonstration of a scientific concept, they can use an educational video game to supplement the lesson and provide students with an interactive experience. In this way, video games can help to fill in the gaps when traditional resources are unavailable, ensuring that students receive a well-rounded education. \cite{molin2017role} However, there’s also existing literature that game-based learning models may be difficult to implement due to resistance and lack of familiarity with its utilization among current teachers along with the lack of technological infrastructure to support such a model. \cite{pan2021implement}(Pan, 2021) But nonetheless, as technology infrastructure improves and educational games become more commonplace in mobile and laptop devices, “the combination of gamification and traditional learning methods can enhance students’ learning outcomes.” \cite{cheung2021application}

In recent years, the spread of misinformation has become a significant issue, with many people being exposed to false or misleading information on a daily basis. To address this problem, some educational video games have been developed to teach players about the importance of fact-checking and critical thinking when it comes to evaluating information. 

One such example is "Bad News," \cite{roozenbeek2019fake} a game developed by the University of Cambridge that simulates the experience of creating and spreading fake news. Players are tasked with building a fake news empire, but they quickly learn that their actions have consequences, as they see the impact their false information has on public opinion. The game teaches players about the techniques used to spread misinformation and provides them with the tools to critically evaluate news and information in the real world. 

Another example is Harmony Square which is a ten-minute online game designed to inoculate people against political misinformation. Developed by a Dutch media collective DROG in collaboration with the U.S. Department of State’s Global Engagement Center and the Department of Homeland Security, the game draws on "inoculation theory" and uses weakened doses of common techniques used in political misinformation to teach players five common manipulation techniques: trolling, emotional language, polarizing audiences, spreading conspiracy theories, and artificially amplifying the reach of content through bots and fake likes. The game’s randomized controlled trial found that people who played Harmony Square were significantly less likely to report sharing misinformation, significantly more confident in their assessment, and found misinformation significantly less reliable after playing. \cite{roozenbeek2020susceptibility}

Fake It To Make It is an online game that puts players in charge of a fake news website and challenges them to create and spread sensationalized news articles to make money. The game is inspired by the Macedonian teens who profited from fake news leading up to the 2016 US election. The game's creator hopes to make players more aware of how and why fake news is written and distributed, encouraging them to be more skeptical of what they encounter in the future. The game is designed to be realistic and could almost be used as a training tool for creating fake news. As players progress through the game, they can expand their fake news empire and earn money by collecting personal information or installing malware. The game's addictive nature and focus on getting more views and money highlights the challenge and enjoyment some fake news creators experience. \cite{plunkett2017videogames}

 In addition to these games, many other educational video games have been developed that address the topic of misinformation, covering subjects such as propaganda, media literacy, and the spread of false information on social media. By incorporating these games into their curriculum, educators can help students to develop the critical thinking skills they need to navigate today's complex media landscape and protect themselves from misinformation.
 
However, a thorough analysis of most of these games shows that they rarely take a more intimate approach to how misinformation on an individual level exploits their biases but instead usually either follow an inoculation theory method or just try to train individuals how to identify fake news off various indicators instead of taking a more holistic approach to the misinformation process. In addition, there aren't really any video games that specifically cover the topic of how AI tools and misinformation are evolving with deep fakes and advanced large learning models like ChatGpt. 

The purpose of this thesis is to propose the creation of a video game that addresses this gap by trying to show how misinformation specifically targets the biases of individuals in order to create a more convincing product and reveal how current misinformation structures operate from start to finish. It will also explore the ways in which AI tools and misinformation have changed and will continue to change the spread of false information. The game will aim to educate players about the potential dangers of AI and its impact on the spread of misinformation and to provide them with the skills and knowledge they need to critically evaluate information in the future. 

The game will use a combination of interactive scenarios and educational content to explore the topic of AI and misinformation. They will be given various case study scenarios of different types of people and how specifically their life stories made them more susceptible to different types of misinformation. Players will also be presented with real-world examples of AI-powered misinformation and will be challenged to identify and dismiss false information. The game will also provide players with an understanding of the ways in which AI can be used to spread false information, including the manipulation of data and the generation of fake news.  

Overall the goal is to provide players with an interactive and engaging way to learn about this important topic and equip them with the critical thinking skills they need to navigate the complex media landscape of the future.

\section{CHAPTER TWO: Game Development and Design}

\subsection{Game Design Overview}
Although the efficacy of debunking and fact-checking in addressing misinformation is a topic of ongoing research and debate, a considerable body of cognitive psychology research underscores the stubborn and lingering influence of misinformation. \cite{nyhan2020taking} Falsehoods can be challenging to correct when encoded in memory and consistent exposure to fake news can heighten the perceived validity of erroneous information. \cite{pennycook2018prior} In response, scholars have increasingly turned to the exploration of more proactive strategies and preventive measures against the spread of misinformation. \cite{roozenbeek2019fake} Given the similarities between the spread of fake news online and the replication of a virus, one promising approach has been to apply the principles of inoculation theory. \cite{kucharski2016study}
Cognitive inoculation follows the same idea as a traditional vaccine where you confer resistance toward a virus by exposing individuals to controlled, weakened doses. \cite{doi:10.1080/03637750802378807} Historically this approach has been used in the misinformation space towards specific topics such as climate change and vaccine usage, however; recently games such as Bad News have been attempting to take this inoculation approach and focus on teaching people to the techniques and intentions behind the creation of misinformation, which has been found to even have efficacy on people who already held deep-seated beliefs on contentious issues. \cite{cook2017neutralizing} Similiarly, In the course of designing the game, I choose to incorporate this inoculation theory by creating the game from the perspective of a misinformation propagandist.
Following the inoculation theory that teaching from the propagandist perspective leads to better learning outcomes, the player is put in controlled scenarios of increasing severity to mimic the dosing schedule of a traditional vaccine. While other games discussed in previous sections are based on a similar concept, what makes my game unique is the emphasis on scenarios that focus on the usage of AI tools and their ability to rapidly create convincing misinformation content rather than just educating the player on more traditional six most common misinformation techniques: impersonating people online, using emotional language, group polarisation, spreading conspiracy theories, discrediting opponents, and trolling. \cite{roozenbeek2019fake} However, the initial game scenario is framed around traditional misinformation concepts in order to give the player a more concrete foundation of the misinformation world to make the scenario about misinformation AI tools a more effective learning experience. \cite{barzilai2014scaffolding}
In addition, research showed that most digital game based learning games such as mine often employ game elements commonly found in commercial games, such as a storyline, feedback, rewards, progress tracking, and challenges.  \cite{6758978} These game components can be mapped to the major game attributes identified by Bedwell et al. (2012), which include rules and goals, assessment, progress, representation, challenge, and safety, all attributes we attempted to incorporate into our game.
The game is broken up into three major sections, a more traditional misinformation scenario where you are a propagandist trying to sway two different types of voters based on the information given to vote for your fictional city council candidate, an AI-focused scenario where you instead now the target of an artificial smear campaign powered by AI chatbots and deep fakes and have to figure out to respond, and a retro section that breaks down in a bit more detail how misinformation works with a few sandbox tools the player can experiment with to improve their understanding. Although the game events are fictional they are modeled off real-world occurrences, potential possibilities, and technologies that already exist today.  
In addition to the player experiencing a central story, the game also has most of its feedback, rewards, and progress tracking mechanisms built into two resources they maintain throughout their gameplay:voter support and money. Throughout the game the player will earn more voter support if they make choices about how they are going to manipulate their voters, however certain actions, like ad creation or buying personal information, will cost money for them to perform. The player can also increase their money with their voter support however through donation ads and other mechanisms. Overall though, the player’s main feedback will be based on how the choices they make affet their voter support score along with tutorial explanations about why certain choices they made cause them to lose or gain voter support. 

\subsubsection{Figure 0: Game Story Board}

\includegraphics[width=\linewidth]{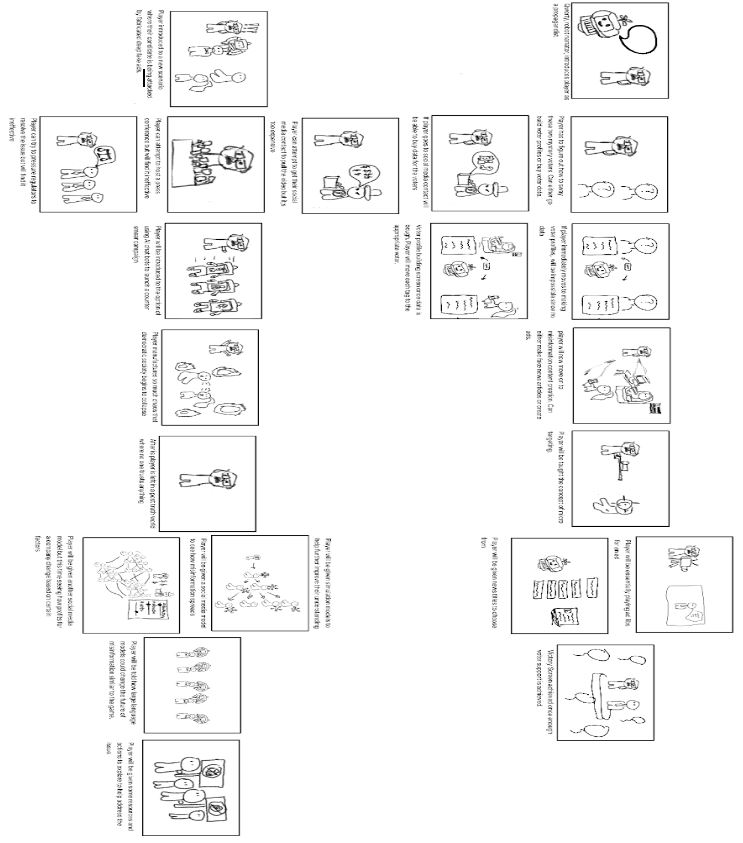}
\label{fig:storyboard}

\subsection{Manipulating Biases, Liberals v.s. Conservatives and personal susceptibility}

\begin{center}
\includegraphics[width=11cm, height=7cm]{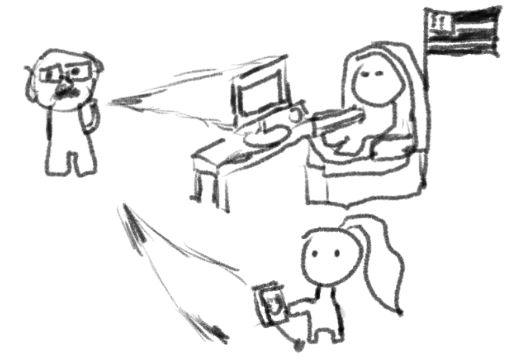}
\label{fig:propagandist}
\end{center}
\subsubsection{Figure 1: Propagandist trying to persuade Bob and Mandy about to support your candidate(Perry)}

In the initial scenario the player is attempting to get city council candidate Perry elected by manipulating two different voters, Bob, who is divorced and an Afghanistan war veteran fighting for custody of his two children, struggling to make ends meet with his job at a local lamp-making plant, and passionate about climate change, and Mandy, who is a passionate liberal who enjoys feminist literature, sustainable agriculture, outdoor activities, plant-based cuisine, attending concerts and exhibitions, writing, and supporting local artists. With this information, the player learns how to create voter profiles based on interest buckets that are derived from their specific interests. The player will be given various tags that describe one of the voters and complete the level by dragging them to the appropriate voter bucket. (Figure 2) If they get an answer wrong, a bubble will pop up explaining why their choice was incorrect.

\begin{center}
\includegraphics[width=11cm, height=7cm]{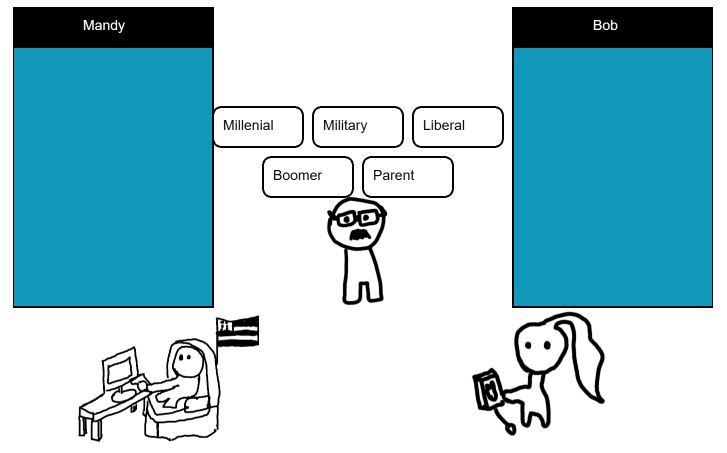}
\label{fig:voterprofiles}
\end{center}
\subsubsection{Figure 2: Building Voter Profiles}

The player will then learn how to generate fake news that targets these biases by being presented with various headlines and ad content choices that would be most effective for the respective voter with an emphasis on emotional appeal.  (Figure 3) Similar to the previous level they will be told if they choose a sub-optimal word prompt along with an explanation.

\begin{center}
\includegraphics[scale=1.15]{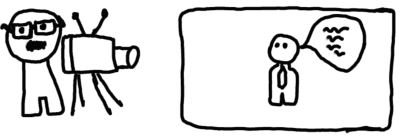}
\label{fig:confab}
\end{center}
\subsubsection{Figure 3: Content Fabrication}

In addition throughout this scenario the player is exposed to ideas about how as a propagandist, the lack of regulation over data privacy allows you to do things like buy personal information from social media and micro-target your content based on things like search history in order to be more effective. (Figure 4)

\begin{center}
\includegraphics[scale=1.15]{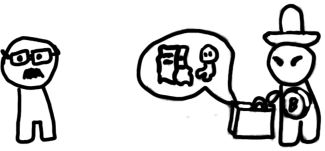}
\label{fig:buyvotes}
\end{center}
\subsubsection{Figure 4: Buying Voter Data}

\subsection{AI Tools and deep fakes}
In the subsequent scenario, after establishing baseline misinformation concepts, the player is then exposed to how AI tools change how easily and rapidly high quality misinformation can be created. The player’s campaign becomes targeted by deep fakes, introducing the concept that people can create convincing fake videos of their political candidate saying things that would alienate their core voters they have been targeting. (Figure 5)

\begin{center}
\includegraphics[scale=1.05]{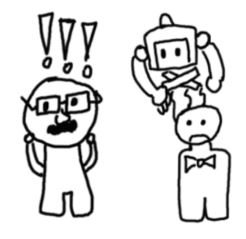}
\label{fig:deepfake}
\end{center}
\subsubsection{Figure 5: Deep Fake Scenario}

The player is then shown how poorly equipped traditional intervention methods are at combating such types of misinformation, with them being presented three options: holding a press conference, contacting their social media company character they have been working with, or leaning on their government contacts to do something about the issue. However, as they explore each solution, they will discover how woefully inadequate current systems that are place are at combating rapidly spread misinformation generated by AI as more deep-faked videos just keep spreading. 
When they attempt to hold a press conference and debunk the video, the effect on their voters base  will not be significant to establish the idea that it’s significantly harder to convince someone that information is false once they’ve established belief in prior information. In addition, they will see how their ability to influence public opinion through a simple press conference is limited as they keep getting new notifications of deep faked videos of their candidate being spread. (Figure 6)

\begin{center}
\includegraphics[scale=.8]{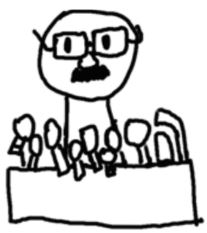}
\label{fig:presscon}
\end{center}
\subsubsection{Figure 6: Press Conference}

When they attempt to pay their social media corporate contact, the player will find the price to take down the video to be exorbitantly high, due to the fact that the social media company website is doing incredibly well with how much controversy and traction the deep-fakes are bringing to their platform. If the user does attempt to get the deep-fakes taken down, they will again see minimal improvements to their voter support bar since these videos had already been shared and watched by millions of people.(Figure 7)

\begin{center}
\includegraphics[scale=.8]{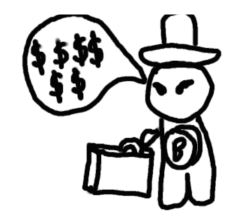}
\label{fig:socialmediaint}
\end{center}
\subsubsection{Figure 7: Social Media Intervention}

The player then can try to choose to lean on their government contacts that they have to try to regulate and take down these deepfakes, but will find that many of the lawmakers have very little experience or knowledge surrounding this technology and that there doesn't exist any laws that regulate or ban this type of content from being created. In addition, the player will be informed that most laws and judicial solutions most likely will take years to pass and not even guaranteed to be effective since it would be akin to outlawing people from photoshopping images with how easy the tools have become available in the fictional world. (Figure 8)

\begin{center}
\includegraphics[scale=.7]{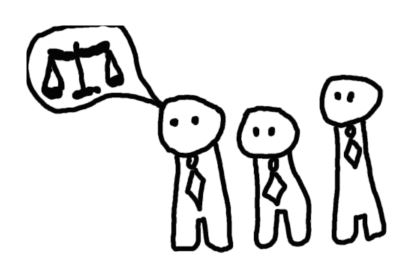}
\label{fig:failreg}
\end{center}
\subsubsection{Figure 8: Failure of Regulation}

However, once the player has exhausted all the traditional options, they will be presented with a nuclear option, where for a relatively cheap amount of money, they can pay some programmers to start running a counter smear campaign using AI social media bots that write realistically sounding anecdotes against their opponent. The idea is to instill the idea into the player that the ability to create and spread misinformation with these tools has been made significantly cheaper and more accessible than traditional methods. In addition, the player will how it’s easier to spread misinformation against a target, than it is to rely on current regulations and companies to reign in the use of these tools without external pressure. (Figure 9)

\begin{center}
\includegraphics[scale=1.1]{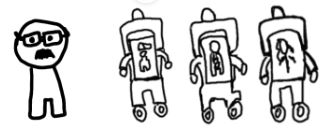}
\label{fig:aiwar}
\end{center}
\subsubsection{Figure 9: AI Chatbot Misinformation Warfare}

At the end of the narrative, the player will have won because they had successfully destroyed the credibility of the opposing candidate. The main consequence they will have though, is that their voter approval bar’s maximum score has actually shrunk due to voter apathy. In addition, any attempt to manipulate voters will be ineffective since many of the voters simply are burnout and don’t know what information they can trust online anymore. (Figure 10)

\begin{center}
\includegraphics[scale=.9]{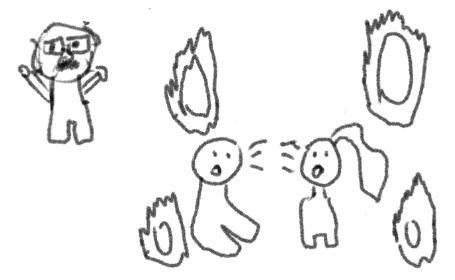}
\label{fig:societalfallout}
\end{center}
\subsubsection{Figure 10: Societal Fallout}

\subsection{Social Media and Government Simulation}
After the narrative concludes and the player has had the opportunity to interact with all these concepts, the game will then present the player with a short lecture blurbs and sandbox tools about everything covered in order reinforce what they have just learned. A lot of research has been done about the efficacy of “activity before content” \cite{suartama2019development} learning, where students are found to be more engaged with a lesson if they have the opportunity to explore the core concepts in a structured activity before a teacher provides the straight content they want the student to learn. \cite{lopez2019examining}
The first tool the player will be given is a simple misinformation propagation model. Misinformation spreads on social media in a way that's similar to how contagious diseases spread in the physical world. \cite{bak2022combining}Just as a disease spreads from one person to another through close contact, misinformation propagates from one individual to the next via social networks, online communities, and message boards, with the interconnectedness among people being harder to control in online spaces compared to say implementing a lockdown where a contagious disease is present. \cite{info:doi/10.2196/24425}To model this spread, I use a network graph that maps the relationships between users, identifying key hubs and nodes where information is most likely to spread rapidly, and simulating the spread of a piece of false information across the network over time. The model can be refined by the player via adjusting the likelihood of individuals sharing information based on their social connections, interests, and other factors.(Figure 11)

\begin{center}
\includegraphics[scale=.9]{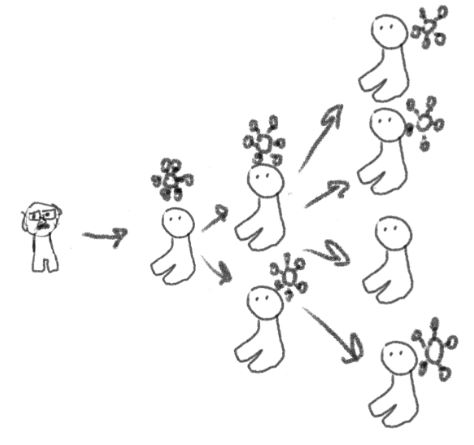}
\label{fig:spreadmodel}
\end{center}
\subsubsection{Figure 11: Misinformation Spread Model}

The next tool the player can experiment with is similar to the previous model, but also introduces how things like bots, lack of moderation, and other factors not only increase the spread of misinformation, but also the profits of the social media company controlling the platform. (Figure 12)

\begin{center}
\includegraphics[scale=.9]{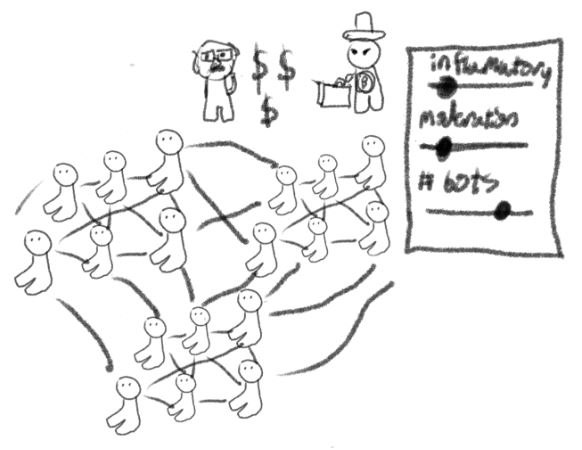}
\label{fig:profit}
\end{center}
\subsubsection{Figure 12: Profitability of Misinformation}

At the end the player is told about how, if left unchecked, the potential dangers of deepfakes and AI chat tools have the potential to pose significant threats to the credibility of online information, with potentially disastrous consequences for society. The player has been exposed to how the use of advanced technologies to manipulate visual and auditory media has enabled the creation of increasingly convincing fake content, which can be used to spread misinformation and propaganda on a massive scale. Similarly, they have been show how the development of AI chat tools has facilitated the spread of disinformation and the creation of persuasive chatbots that can be used to manipulate online conversations and influence public opinion. (Figure 13)

\begin{center}
\includegraphics[scale=.6]{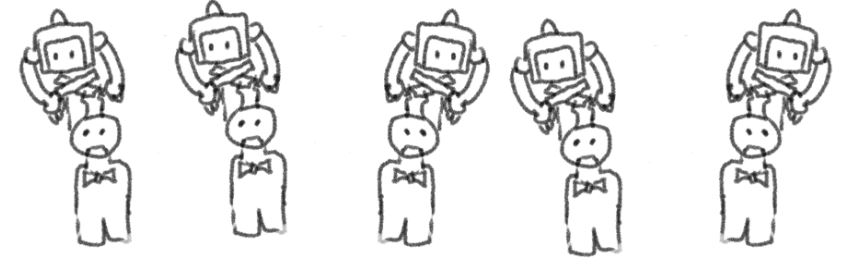}
\label{fig:badfuture}
\end{center}
\subsubsection{Figure 13: The potentially disastrous future}

\subsection{Call to Action}
The purpose of the game is then stated for them as not only to expose them to these concepts so that they are less susceptible to such techniques in the future, but also teach them how important it is for them to advocate for researchers and policymakers to prioritize the development of effective strategies for detecting and mitigating the risks posed by deepfakes and AI chat tools, in order to safeguard the integrity of online information and protect democratic societies from the threat of large-scale deception and manipulation. (Figure 14) The player is also given a closing message about how the power of collective organizing and staying informed can play a crucial role in mitigating the issue of misinformation spread through deepfakes and AI chat tools. By them staying informed about the latest developments in these technologies, they as individuals can be better equipped to identify and flag suspicious content or interactions that may be spreading misinformation. Additionally, they are told how by coming together as a community, individuals can use their collective voices to push for greater transparency and accountability from tech companies and policymakers. Finally, they are told about how they can advocate for the development of tools and technologies that can help detect and combat deepfakes and AI chatbots, as well as supporting policies and regulations that prioritize the protection of online information and individual privacy. “Ultimately, by staying informed and working together, we can help ensure that the internet remains a trustworthy source of information and a tool for promoting democratic values and social progress.”

\begin{center}
\includegraphics[scale=.6]{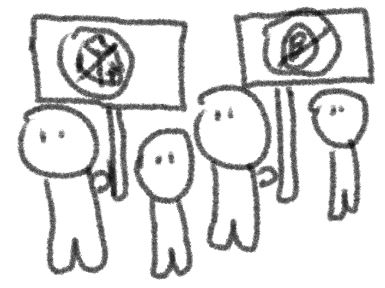}
\label{fig:advandorg}
\end{center}
\subsubsection{Figure 14: Advocacy and Organizing}

\section{CHAPTER THREE: Game Development}
\subsection{Game Host}
The game is a standard javascript web application and is hosted via Netifly. It can be played at https://lucky-phoenix-9ce30d.netlify.app/ and all code used can be found at  https://github.com/WillieShi/Fabricating\_Reality .

\subsection{Game Technology}
In the development of the educational misinformation video game, various technologies were utilized to create an immersive and engaging user experience. One of the key technologies used in the game was PIXI.js, a powerful 2D rendering engine that allowed for the creation of high-quality graphics and animations. With PIXI.js, the game was able to deliver a visually appealing and responsive interface that enhanced the user's overall experience. 
To complement the graphics, Howler.js was utilized for the game's audio components. Howler.js is a powerful audio library that enables the integration of high-quality sounds and music into the game. This not only enhanced the overall user experience but also helped to create a more immersive environment for the user. 
Another key technology used in the development of the game was Tween.js. This library was utilized for tweening animation effects in the game. With Tween.js, the developers were able to create smooth and seamless animations that added a level of polish to the overall user experience.
 In addition, Balloon.css was used for pop-up tooltips, providing a more engaging and interactive learning experience for the user. The tooltips were an effective way to convey important information about the game and its mechanics, as well as the misinformation topics covered in the game. 
To manage asynchronous code and simplify complex logic, Q was utilized for promises. This library made it easier to handle asynchronous events and streamline the development process. Moreover, MinPubSub was used for publish/subscribe, a lightweight event system that helped to decouple the game's components and improve code reusability. 
Finally, Pegasus was used in the game's development, allowing the developers to retrieve data from external APIs with ease. With Pegasus, the team was able to quickly and efficiently integrate external data into the game, reducing development time and effort. Overall, the use of these technologies played a critical role in the development of the educational misinformation video game, enabling the creation of an engaging, immersive, and educational experience for the user.

\subsection{Usability Feedback}
As part of the development process for the educational misinformation video game, usability testing was conducted with a small sample of roughly a dozen users to ensure that the game was user-friendly, intuitive, and engaging. The usability testing involved gathering feedback from a sample of potential users who played the game and provided feedback on their experience.
To conduct the usability testing, a sample of participants was recruited and provided with a set of tasks to complete while playing the game. The tasks were designed to test different aspects of the game's interface and gameplay mechanics, such as navigating menus, interacting with characters, and completing challenges related to misinformation.
During the testing, participants were encouraged to think aloud and share their thoughts and feedback on the game as they played. The feedback collected during the testing was then compiled and analyzed to identify areas of the game that needed improvement.
Based on the feedback received, several changes were made to the game's interface and gameplay mechanics to improve the user experience. For example, some participants reported difficulty understanding the game's objectives or mechanics, which led to changes in the tutorial system and game instructions. Additionally, feedback was gathered on the game's visual and audio components, leading to adjustments to the graphics and sound effects. One example of this is when a user commented it was unclear that players can go back to previous levels to review information so a small tutorial explanation and system reminder was added for clarification. (Figure 14)

\begin{center}
\includegraphics[scale=.7]{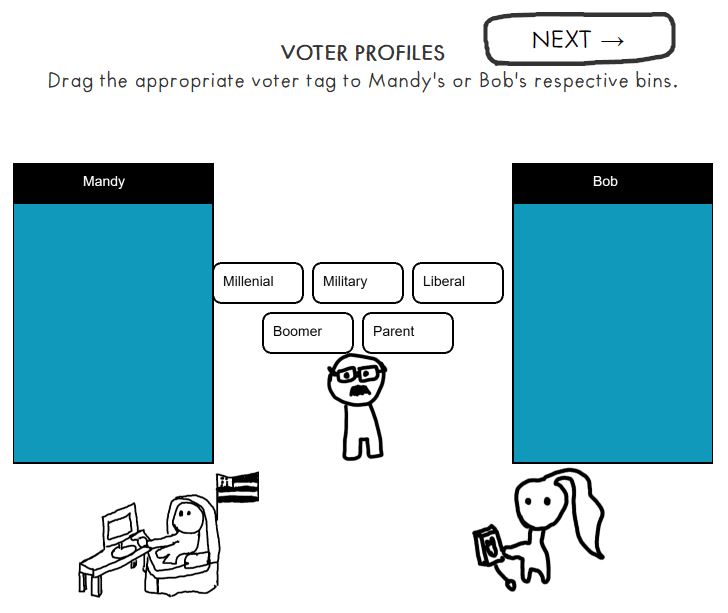}
\label{fig:usage1}
\end{center}
\subsubsection{Figure 15: Tutorial Text Added}

Another major concern that was raised was making sure that users without a technology background were able to grasp any technical jargon presented in the game, since it’s marketed towards groups who traditionally don’t have access to a quality education about computer science and the internet. In order to address these concerns, an explanation bot character named Qwerty was added that could be clicked in order to give explanations on certain concepts the user may not be familiar with. (Figure 15)

\includegraphics[width=\linewidth]{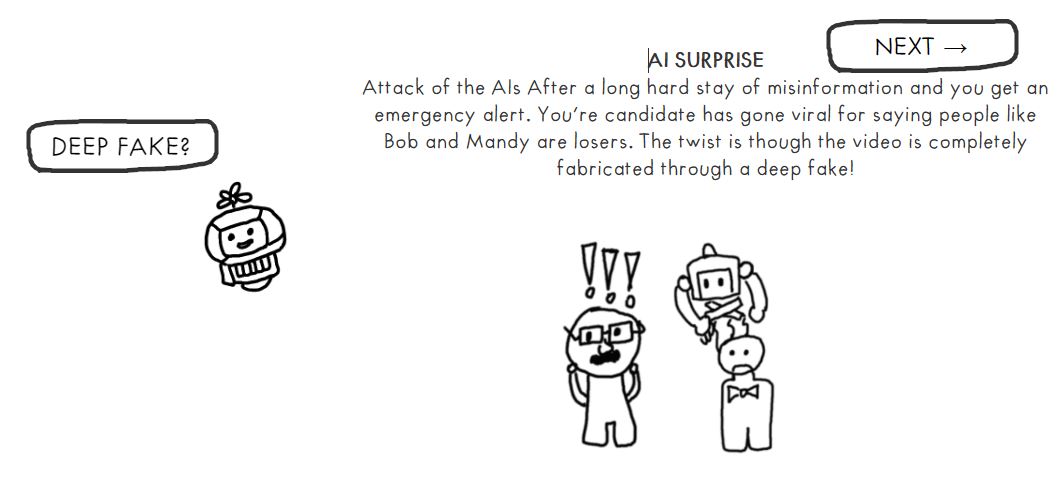}
\label{fig:usage2}
\includegraphics[width=\linewidth]{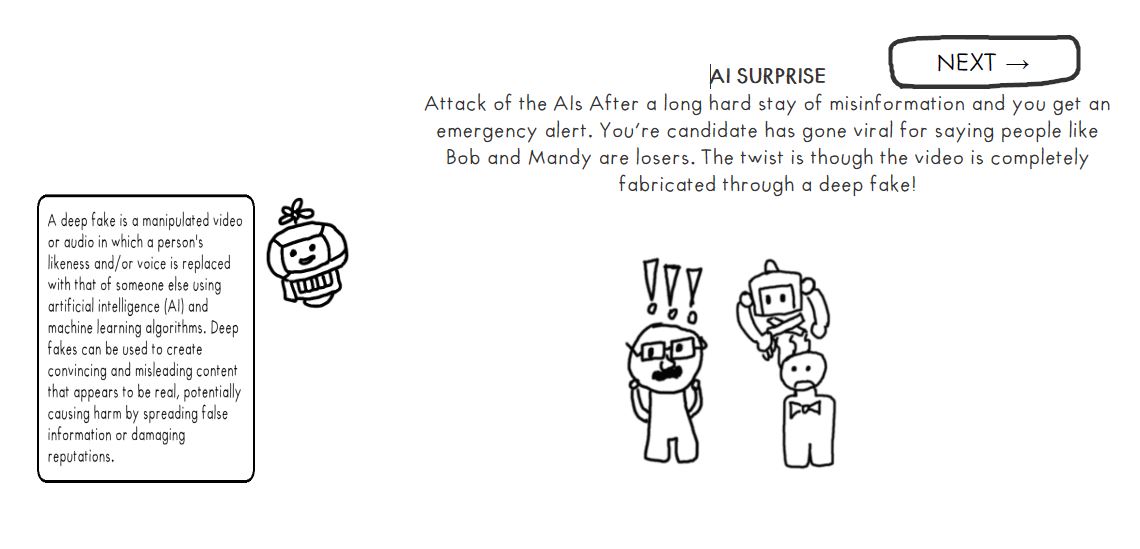}
\label{fig:usage3}
\subsubsection{Figure 16: Qwerty Tool Tip Added}

Overall, the usability testing provided valuable feedback on the game's strengths and weaknesses and helped to identify areas for improvement. By incorporating the feedback received during testing, the game was able to provide a more engaging and intuitive experience for its users, enhancing its educational value and impact.

\subsection{Future Use Cases}
While the development of the educational misinformation video game was a significant step towards improving educational outcomes around misinformation, there is a need for further testing to determine the effectiveness of the game in achieving its intended goals. Unfortunately, due to time and resource constraints, it was not possible to conduct extensive testing on the game during its development phase.
However, I believe that this tool can be used in conjunction with a pre and post-assessment survey to evaluate its effectiveness in improving educational outcomes around misinformation. This approach would involve administering a survey to participants before and after playing the game to measure changes in their knowledge, attitudes, and behaviors regarding misinformation.
By comparing the pre and post-assessment survey results, it would be possible to determine the extent to which the game was effective in achieving its intended outcomes. This information could then be used to improve the game and develop more effective strategies for educating individuals about misinformation.
Furthermore, it is my hope that other researchers and educators will utilize this tool to further evaluate its effectiveness in improving educational outcomes around misinformation. As more data is gathered through research studies, new games can be created based on this one’s shortcomings to become an even more effective tool for educating individuals about misinformation.
 In conclusion, while the testing of the educational misinformation video game was limited during its development phase, the potential for this tool to improve educational outcomes around misinformation is significant. By using pre- and post-assessment surveys to evaluate the game's effectiveness, we can better understand its impact on individuals' knowledge, attitudes, and behaviors toward misinformation. I hope that this tool will continue to be developed and improved upon by researchers and educators to create a more informed and educated society.

\section{Conclusion}
This thesis highlights the critical need to address the dangers of misinformation in today's society, given its potential to cause significant harm to individuals and society as a whole. The proliferation of false information through various mediums has made it increasingly difficult for people to separate truth from fiction, and this has serious consequences, such as spreading fear, division, and even undermining democracy.
It is important to recognize that social media companies, governments, and traditional educational institutions have a responsibility to address the issue of misinformation. However, their incentives are often misaligned, making it difficult for them to be effective solutions on their own. Social media companies may prioritize profit over accuracy, while governments may be hesitant to regulate information sources for fear of impinging on free speech, and educational institutions may not have the necessary resources or expertise to address the issue adequately.
As deepfake and large language model (LLM) technologies become more available, society is facing an even greater challenge in combating the spread of misinformation. These technologies allow for the creation of highly realistic fake images, videos, and text that can be difficult to distinguish from genuine content.
Unfortunately, society is currently woefully unprepared to deal with the proliferation of misinformation brought about by these technologies. Most individuals lack the skills and knowledge necessary to assess the credibility of digital content, especially in the face of highly sophisticated and convincing fake content.
In response, this thesis explored developing an educational game as an unique solution to address the issue of misinformation. By providing a safe and controlled environment for players to practice critical thinking skills and develop an understanding of the tactics used to spread false information, these games can help fill the information gap that exists in internet education. They can also promote awareness about the importance of verifying information before sharing it, which can ultimately help reduce the spread of misinformation. In addition, by incorporating modules on deepfakes, LLMs, and other emerging technologies, these games can help individuals develop the skills necessary to distinguish between genuine and fake content, but also make individuals aware how these technologies are manipulating information spaces.
By doing so, it is hoped that the game will help reduce the spread of misinformation and promote a more informed and discerning public.
While the game is still in production, it is expected that it will address some of the shortcomings of previous games in the space, such as failing to adequately address how modern technologies like chatbots and deep fakes have made individuals even more vulnerable to misinformation. The game code is publicly available for contributors to suggest changes or outright create their own product since it’s distributed under a Creative Commons Zero license making it fully available for public domain use.
As this work demonstrates, there is a need for continued research and development in the field of misinformation games, and it is hoped that others will be inspired by the importance of this work and either improve on the existing code for this game or create their own around the AI Misinformation Space. By doing so, we can make significant strides in addressing the issue of misinformation and promoting a more informed and responsible public.

\section*{Acknowledgments}
 I thank my main thesis advisor Dr. Dhiraj Murthy and my second reader Dr. Min Kyung Lee for all the academic support and advice they have given me throughout writing this thesis. I would not have been able to accomplish this without their expertise and guidance. I thank Shaurya Pathania for his help formatting plots, editing, and migrating this manuscript to LaTex.

\bibliographystyle{unsrt}  
\bibliography{references}

\end{document}